\documentclass[aps,preprint,longbibliography]{revtex4-1}%
\usepackage{amsfonts}
\usepackage{amsmath}
\usepackage{amssymb}
\usepackage{graphicx}%
\providecommand{\U}[1]{\protect\rule{.1in}{.1in}}

\begin{document}

\title{Visco- and plasto-elastic fracture of nano-porous polymer sheets}
\author{Takako Tomizawa and Ko Okumura\\Department of Physics, Ochanomizu University, Tokyo, Japan}

\begin{abstract}
We study the dependence of the fracture surface energy on the pulling velocity
for nano-porous polypropylene (PP) sheets to find two components: the static
and dynamic ones. We show that these terms can be interpreted respectively as
plastoelastic and viscoelastic components, as has been shown for soft
polyethylene (PE) foams in a previous work. Considering significant
differences in the pore size, volume fraction and Young's modulus of the
present PP and previous PE sheets, the present results suggest a universal
physical mechanism for fracture of porous polymer sheets. The simple physical
interpretation emerging from the mechanism could be useful for developing
tough polymers. Equivalence of Griffith's energy balance in fracture mechanics
to a stress criterion is also discussed and demonstrated using the present
experimental data.

\end{abstract}
\date{\today}
\maketitle

\section{Introduction}

Cellular solids, which are porous materials with a well-defined pore size, are
found in natural and artificial materials and are a useful form of materials
\cite{CellularSolids}. Cork, balsa, and apples \cite{Vincent1996Apple} are
cellular solids originated from plants, and the stereom of echinoderms
\cite{Emlet1982BioBull}, skeleton of a certain sponge
\cite{Aizenberg2005Science}, and frustle of diatoms \cite{DiatomNature2003}
are examples from creatures. Such materials possess mechanical advantages
because they can be light, strong, shock-absorbing, and heat-retaining.
Accordingly, active studies have been performed on mechanical and fracture
mechanical properties, focusing on an important parameter for cellular solids,
the volume fraction of the matrix material $\phi$ \cite{CellularSolids}.
However, studies on velocity dependent properties of their fracture are
relatively limited, compared with intensive studies that have been performed
on other materials such as adhesive
\cite{GentPeelingRate1972,Chaudhury1999Rate,morishita2008contact,Creton2002Adhesion,bhuyan2013crack}%
, laminar \cite{Kinloch1994peelingRate}, viscoelastic
\cite{GreenwoodJohnsonRate,schapery1975theory}, weakly cross-linked
\cite{PGGtrumpet,saulnier2004adhesion}, biopolymer gel \cite{lefranc2014mode},
and biological composite \cite{bouchbinder2011viscoelastic} materials,
including recent active experimental
\cite{Tsunoda2000,MoridhitaUrayama2016PRE,morishita2017crack}, numerical
\cite{kubo2017velocity,aoyanagi2017}, and theoretical
\cite{sakumichi2017exactly,Okumura2018} studies on the velocity jump in crack
propagation in elastomers.

Previously, we studied mechanical and fracture mechanical properties of soft
solidified foam of non-cross-linked polyethylene. Young's modulus $E$, the
characteristic pore size $d_{0}$, and the volume fraction $\phi$ of the foams
were of the orders of 1 MPa, 1 mm, and 0.03, respectively. For the soft foams,
we established scaling laws for Young's modulus and the fracture surface
energy as a function of $\phi$. The scaling laws thus found are different from
the ones established for well-studied hard cellular solids of Young's modulus
typically around 3000 MPa \cite{Shiina06}. Furthermore, we revealed for the
same soft solidified foams a simple relation between the fracture surface
energy (required at the crack initiation) and pulling velocity with a clear
physical interpretation \cite{Kashima2014}. Here, we test this simple
description for the velocity dependence of the fracture energy, using a porous
polymer of different nature. The present ample is not polyethylene (PE) but
polypropylene (PP) and $E$, $d$, and $\phi$ in the present case are
significantly different from those in the previous study: $E\sim200$ MPa,
$d_{0}$ $\sim1$ $\mu$m, and $\phi\sim0.5$. As a result, we find that the same
description is well-applicable for this quite different material and suggest
that the simple description proposed in the previous study can be universally
relevant to a certain class of porous polymers.

\section{Experimental Section}

\subsection{Materials.}

\begin{figure}[h]
\begin{center}
\includegraphics[width=0.5\textwidth]{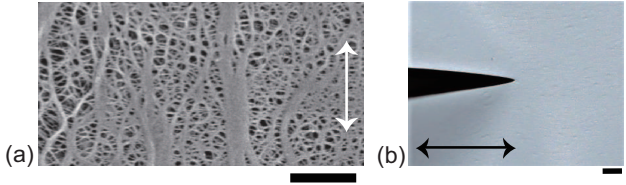}
\end{center}
\caption{(a) A SEM image of the porous polypropylene sheet used in the
experiment (Copyright 2018 by Mitsubishi Chemical Corporation). The length of
the scale bar is 1 $\mu$m. (b) Magnified view near a crack tip. The machine
direction are shown by double-headed arrows, which is perpendicular to the
pulling direction. The length of the scale bar is 1 mm.}%
\label{Fig1}%
\end{figure}

In this experiment, we use sheets of polypropylene (PP) that possesses a
porous structure as shown in Fig. \ref{Fig1}(a). The structure may be
characterized by a few different length scales and the largest scale is around
1 $\mu$m, which could be comparable to 10$^{-5}$ m$,$ as seen from the SEM
image. The volume fraction $\phi$ and the thickness of the sheet are $0.44$
and 23 $\mu$m, respectively. The sheet is fabricated from bulk with a
stretching process and thus tends to be stronger in the direction, which will
be called the machine direction.

\subsection{Mechanical measurement}

The elongation of a sheet from the natural length and the tensile force acting
on the sheet were measured with a hand-made setup, which was used in our
previous study \cite{Kashima2014}. This setup is equipped with horizontally
placed two pairs of clamp bars specially designed for sheet samples of
dimension comparable to 50 cm to avoid any slips at the clamps and local
slacking of a sheet sample under stretch. The horizontal width of the sample
is 10 cm for failure stress measurements (5 cm for force-extension
measurements) and the vertical height, i.e., the distance between the inner
edges of the clamp bars is 12.5 cm. The bottom pairs of clamp bars are fixed
to the setup frame, while the position and speed of the mobile upper pairs of
clamp bars can be controlled by a slider system (EZSM6D040 K, Oriental Motor)
through a digital force gauge (FCC-50B, NIDEC-Shimpo). Each measurement is
performed for a given fixed velocity of the upper clamp, which defines the
pulling velocity $V$. All the experiments are performed with setting the
machine direction perpendicular to the vertical direction, in which direction
the constant stretching speed $V$ is given by the controlled slider system.
The pulling velocity $V$ is varied from 0.03 mm/sec to 0.4 mm/sec.

Fracture mechanical measurements are performed with the same setup but with
introducing a macroscopic line crack with a sharp knife at the center of the
sample (see Fig. \ref{Fig1}(b)). The line crack is created in the horizontal
direction (i.e., the machine direction) and the length $2a$ is varied from 2
to 32 mm.

\section{Results}

\subsection{Stress-strain curve and Young's modulus}

\begin{figure}[h]
\begin{center}
\includegraphics[width=\textwidth]{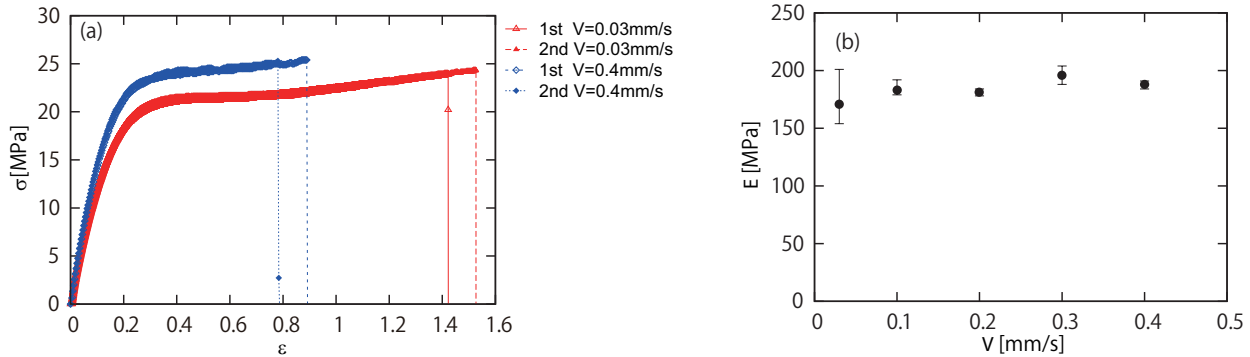}
\end{center}
\caption{(a) Stress-strain curves obtained from the force-extension
measurements performed with samples without any macroscopic cracks at
extension velocities, $V=0.03$ and $0.4$ mm/sec. (b) Young's modulus $E$
determined in the initial regime in the stress-strain curve as a function of
the stretching velocity $V$. }%
\label{Fig2}%
\end{figure}

Figure \ref{Fig2}(a) shows typical results for the relation between stress
$\sigma$ and strain $\varepsilon$ of the sheet sample at different velocities.
Nearly overlapped two data sets at each velocity show a reasonable
reproducibility of the experiment. The relation shows a weak nonlinearity with
a weak dependence of velocity. However, in the initial regime in which
$\varepsilon$ and $\sigma$ are less than $\sim0.1$ and $\sim2 $ MPa,
respectively, the relation is almost linear and is almost independent of
velocity. In fact, Young's modulus extracted from this region is practically
constant as shown in Fig. \ref{Fig2}(b). Here, the modulus at each velocity is
determined as the average of the three moduli obtained from the initial
regimes of three sets of force-elongation measurements performed at each
velocity, with the bottom and top values of each error bar showing the minimum
and maximum of the three results.

\subsection{Failure stress with a line crack and fracture surface energy}

\begin{figure}[h]
\begin{center}
\includegraphics[width=0.35\textwidth]{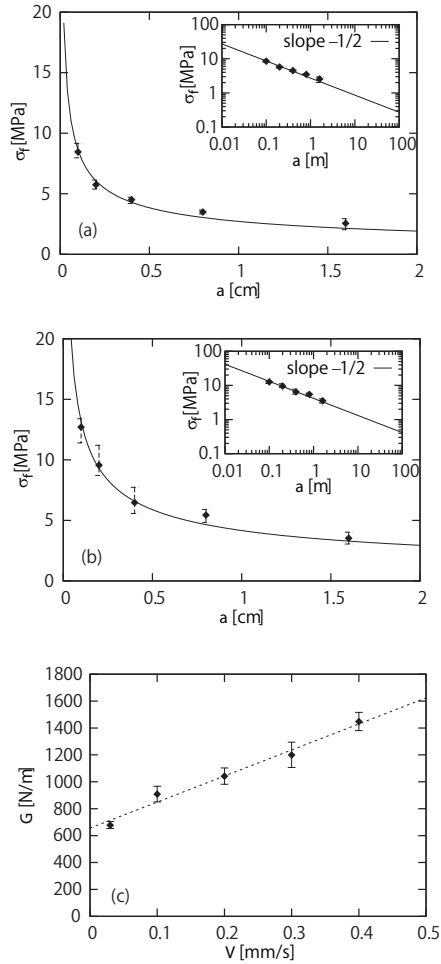}
\end{center}
\caption{Failure stress $\sigma_{f}$ as a function of crack size $2a$ at
$V=0.03$ mm/sec (a) and $V=0.4$ mm/sec (b). The insets show corresponding
plots on a log-log scale. (c) Fracture surface energy $G$ as a function of
velocity $V$.}%
\label{Fig3}%
\end{figure}

Figure \ref{Fig3}(a) and (b) show typical results for the relation between the
failure stress $\sigma_{f}$ and the half length of the line crack $a$. The
measurement of $\sigma_{f}$ for given $a$ and $V$ are performed three times
and the average of the three values is plotted with an error bar in the
figure. When data are plotted on a log-log scale, the data for a given $V $
collapse clearly on a straight line with slope $-1/2$. This justifies that we
use the following Griffith's formula to estimate the fracture surface energy
$G$ for a given $V$ \cite{Anderson}:%
\begin{equation}
\sigma_{f}=\left(  \frac{2EG}{\pi a}\right)  ^{1/2} \label{eq1}%
\end{equation}

Figure \ref{Fig3}(c) shows the relation between the pulling velocity $V$ and
the fracture surface energy $G$ on the basis of Eq. (\ref{eq1}). The relation
can be well described by the linear form: $G=A+BV$. Here, $A$ and $B$ are
constants independent of $V$. On the other hand, it is theoretically justified
and experimentally confirmed that $G$ is proportional to the volume fraction
$\phi$ \cite{Shiina06,Kashima2014}. Exploiting this property, we rewrite the
expression $G=A+BV$ in the following form:%

\begin{equation}
G=\phi G_{0}(1+V/V_{0})
\end{equation}
with introducing $G_{0}$ and $V_{0}$, which are independent of $V$ ($G_{0}$
and $V_{0}$ are defined by the relations $A=\phi G_{0}$ and $B=\phi
G_{0}/V_{0}$).

In the following, we justify that the above expression can be reasonably well
interpreted as in the following form:%
\begin{equation}
G=\phi(\sigma_{Y}+\eta(V/d))\delta\label{eq3}%
\end{equation}
We examine the validity of the above expression at the level of scaling laws.
For simplicity we regard $\phi\sim1$ and ignore this factor in the following.
The first term $\phi\sigma_{Y}\delta\sim$ $\sigma_{Y}\delta$ is considered as
the standard expression for plastic fracture with $\sigma_{Y}$ the yield
stress and $\delta$ is the crack opening distance \cite{Anderson}. Considering
a typical value of $\sigma_{Y}$ for PP, say, 50 MPa, the opening distance
$\delta$ can be estimated as $\sim10^{-5}$ m\ because $\phi G_{0}\sim G_{0}$
($\sim\sigma_{Y}\delta$) is estimated as 650 J/m$^{2}$ from Figure
\ref{Fig3}(c). This value of $\delta$ is comparable to the largest
characteristic scale of the porous structure in the sample; it is quite
natural that we expect that crack opening distance scale as this length scale.
The second term, in particular the quantity $\eta(V/d)$, just describe the
viscous stress that we have to add to $\sigma_{Y}$ in the dynamic case. Here,
$d$ describes the length scale around the crack tip dynamically affected and
it is natural to assume that this scale as $\delta$. Taking the viscous effect
into account is reasonable because the glass transition temperature of PP is
well below ambient temperature (typically 0 $^{\circ}C$), at which the
experiments were performed and the viscosity $\eta$ is estimated as
2$\times10^{6}$ Pa.s from Figure \ref{Fig3}(c) by estimating the value $\phi
G_{0}/V_{0}\sim G_{0}/V_{0}$, which scale as $\eta\delta/d\sim\eta$. To gain
physical insight, we can roughly estimate the number of monomers $N$ in the
entangled polymer by invoking the reptation model \cite{PGG1979scaling}. This
crude estimate gives $\eta\sim\eta_{0}N^{3}/N_{e}^{2}$ with $N_{e}$ the
entangle distance ($\sim100$) and $\eta_{0}$ the viscosity of monomers
($\sim1$ mPa.s). As a result, we obtain a plausible value, $N\sim10^{4}$.
Since we have established the same reasoning for Eq. (\ref{eq3}) in
Ref.\cite{Kashima2014} for soft foam sheets of polyethylene, the present
result suggests a certain degree of universality of Eq. (\ref{eq3}). Our
result suggests that the dynamic toughness is significantly increased if the
number of monomers $N$ is increased, because the velocity dependent term in
$G$ contains the viscosity $\eta$ and, even if the reptation model is not
appropriate, $\eta$ is strongly dependent on $N$.

Equation (\ref{eq3}) established as above can be interpreted as composed of
the static term reflecting a plastoelastic effect and the dynamic term
reflecting a viscoelastic effect, as explained as follows. This expression is
composed of two terms, one independent of $V$ and the other dependent on $V$.
In other words, $G$ is composed of the static and dynamic parts. The static
part reflects a plastic effect because it is proportional to the yield stress
$\sigma_{Y}$. The dynamic part corresponds to viscous effect because it
originates from viscous stress. On the other hand, $G$ in the present study is
determined by linear-elastic fracture mechanics. Therefore, it is natural to
interpret the static and dynamic terms as plastoelastic and viscoelastic
effects, respectively.

\section{Equivalence of Griffith energy balance to a stress criterion}

As we discussed in our previous paper, the energy balance is equivalent to a
stress criterion, if we note that the stress concentration is cutoff at the
length scale below which the continuum description fails, although Griffith's
energy balance is sometimes distinguished from the stress criteria for
fracture. (This length scale corresponds to the largest scale characterizing
the porous structure and, thus, will be called $d$ in the following.) This can
be experimentally confirmed for the present material. To see this, let us
briefly review how the energy balance reduces to a stress criterion. We
introduce a critical failure stress $\sigma_{c}$ for a given material through
the following relation:%
\begin{equation}
\sigma_{c}\simeq(EG/d)^{1/2} \label{eq4}%
\end{equation}
This comes from the well-known Griffith's formula for the failure stress
$\sigma_{f}\simeq(EG/a)^{1/2}$ for a crack of length $\sim a$ and the idea of
Griffith's cavities. We further introduce the maximum stress that appears at
the crack tip at the critical of failure at which the remote stress
$\sigma_{0}$ is equal to the failure stress $\sigma_{f}$:%
\begin{equation}
\sigma_{m}\simeq\sigma_{f}(a/d)^{1/2} \label{eq5}%
\end{equation}
This comes from the well-known Inglis' stress concentration formula for the
stress distribution around a crack $\sigma(r)\simeq\sigma_{0}(a/r)^{1/2}$ at a
distance $r$ from the crack tip, which should be cutoff at the scale $d$.
(This relation $\sigma(r)\simeq\sigma_{0}(a/r)^{1/2}$ has been confirmed in
previous numerical studies \cite{Nakagawa,Aoyanagi2009JPSJ}.) The stress
criteria that is equivalent to Griffith's energy balance is then given by%
\begin{equation}
\sigma_{c}\simeq\sigma_{m} \label{eq6}%
\end{equation}
In fact, from Eqs. (\ref{eq4}) to (\ref{eq6}), we recover Griffith's formula.

\begin{figure}[h]
\begin{center}
\includegraphics[width=0.35\textwidth]{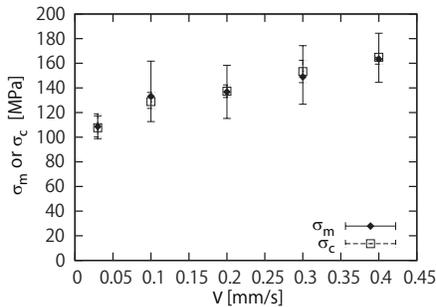}
\end{center}
\caption{$\sigma_{c}$ vs $\sigma_{m}$.}%
\label{Fig4}%
\end{figure}

To confirm the above description is relevant to the present experiment, we
experimentally determine $\sigma_{c}$ and $\sigma_{m}$, respectively, from Eq.
(\ref{eq4}) (using measured values of $E$, $G$, and $d$) and from Eq.
(\ref{eq5}) (using measured values of $\sigma_{f}$, $a$, and $d$). We here set
the coefficient for Eq. (\ref{eq4}) to be 1 and that for Eq. (\ref{eq5}) to be
0.841 and use $d=10^{-5}$ m for both equations, for simplicity. (If we denote
dimensionless numerical coefficients suppressed in Eq. (\ref{eq4}) to Eq.
(\ref{eq6}) as $c_{4}$, $c_{5}$, and $c_{6}$, respectively, Eq. (\ref{eq6})
can be expressed as $(EG/d)^{1/2}=(c_{6}c_{5}/c_{4})\sigma_{f}(a/d)^{1/2}$,
which implies $c_{6}c_{5}/c_{4}$ is close to 0.841.) The results are shown in
Fig. \ref{Fig4}, which confirms that the stress criteria Eq. (\ref{eq6}) is
satisfied for the experimentally observed fractures and, thus, the above
description is consistent with our experimental results.

\section{Conclusion}

We have studied the fracture surface energy at the onset of crack initiation
as a function of the pulling velocity. As a result, we find that the fracture
energy is described by the static and dynamic components. The former
corresponds to the standard plastic fracture. The latter reflects the viscous
flow that occurs at the crack tip. The both components are characterized by
the length scale that is comparable to the pore size. The existence of the
dynamic and viscous component suggests a simple principle for toughening: the
increase in the degree of polymerization could greatly enhance the dynamic
toughness. The equivalence of Griffith's energy balance to a stress criterion
is also demonstrated. These results are completely in parallel with the
results obtained for significantly different porous polymer sheets, which
suggests the universality of the present results although further studies will
be needed to clarify the limitation of the present interpretation.

Considering that the basic physical properties of polymer materials are quite
dependent on preparation crystallinity, molecular orientation, size and shape
of porous structure, types of polymers \cite{strobl1997physics}, the emergence
of this kind of universality may be unexpected. (The fracture of crystalline
polymers on microscopic scales are started to be reproduced in simulations
\cite{higuchi2018fracture}.) We consider that one of the reasons for this
universality comes from that we introduce a macroscopic crack, whose scale is
much larger the characteristic scales for such as crystallinity, domains in
which molecules are oriented and porous structure. On such a macroscopic
scale, it is known that polymers universally exhibit yielding and viscous
flowing in a similar manner. When these points are considered, it would not be
so surprising if the fracture associated with a macroscopic crack has
universally governed by yielding and viscous flowing in their simplest forms,
which we are suggesting.

The study concerns the fracture surface energy required at the onset of crack
initiation. Such a fracture energy should be in general distinguished from the
fracture energy required during crack propagation. In fact, recently, we have
studied the relation between the energy release rate, which can be interpreted
as the fracture surface energy required during crack propagation, and the
crack-propagation velocity, for the same material studied in the present study
\cite{Takei2018}. As a result, we did not observe crack propagation in the
velocity range studied in the present study. However, in the previous study
\cite{Takei2018}, from a technical reason, the constant-speed crack
propagation with a fixed-grip condition was initiated some time after we set a
given strain to samples and we consider that the effect of stress relaxation
due to this preparation time could tend to suppress crack propagation. This
point will be further discussed elsewhere.

\section*{Acknowledgements}

The authors are grateful to Mitsubishi Chemical Corporation for providing
nano-porous sheet samples and the SEM image shown in Fig. \ref{Fig1}(a). The
authors thank Dr. Atsushi Takei for technically helping them for experiments.
This work was partly supported by ImPACT Program of Council for Science,
Technology and Innovation (Cabinet Office, Government of Japan).

\bibliographystyle{unsrt}
\bibliography{C:/Users/okumura/Documents/main/JabRef/granular,C:/Users/okumura/Documents/main/JabRef/fracture,C:/Users/okumura/Documents/main/JabRef/wetting}

\end{document}